\documentclass[]{article}

\usepackage[centertags]{amsmath}
\usepackage{amsfonts}

\newcommand{\Di}{\Delta_m}
\newcommand{\Dj}{\Delta_n} 
\newcommand{\Dim}{\Delta_{-m}}
\newcommand{\Djm}{\Delta_{-n}}

\newcommand{\ioo}{_{m,n}}
\newcommand{\ii}{_{m+1,n}}
\newcommand{\jj}{_{m,n+1}}
\newcommand{\ij}{_{m+1,n+1}}
\newcommand{\iii}{_{m+2,n}}
\newcommand{\jjj}{_{m,n+2}}
\newcommand{\iim}{_{m-1,n}}
\newcommand{\jjm}{_{m,n-1}}
\newcommand{\iimim}{_{m-2,n}}
\newcommand{\jjmjm}{_{m,n-2}}
\newcommand{\jimjm}{_{m-1,n-1}}
\newcommand{\jimj}{_{m-1,n+1}}
\newcommand{\jijm}{_{m+1,n-1}}
\newcommand{\iiijm}{_{m+2,n-1}}
\newcommand{\iimjj}{_{m-1,n+2}}
\newcommand{\iimimj}{_{m-2,n+1}}
\newcommand{\iijmjm}{_{m+1,n-2}}

\def\mref#1{(\ref{#1})}

\newtheorem{conc}{Conclusion}
\newtheorem{thm}{Theorem}

\begin{document}

\title{\bf Darboux transformations for a 6-point scheme
\thanks{The initial stage of the work
was supported by Polish KBN grants 2 PO3B 126 22 and
1 P03B 017 28 while at the final stage (starting from 1st April 2005) the paper was supported solely
 by the European Community under a  Marie Curie Intra-European Fellowship, 
 contract no MEIF-CT-2005-011228.}}


\author{M. Nieszporski
\thanks{Department of Applied Mathematics, University of Leeds, Leeds LS2 9JT, UK
e-mail: maciejun@maths.leeds.ac.uk, tel: +44 113 343 5149 fax: +44 113 343 5090}
\thanks{Katedra Metod Matematycznych Fizyki, Uniwersytet Warszawski ul. Ho\.za 74, 00-682 Warszawa, Poland
e-mail: maciejun@fuw.edu.pl, tel: +48 22 621 77 57, fax: +48 22 622 45 08}}

\maketitle

\begin{abstract}
We introduce (binary) Darboux transformation 
for general differential equation
of the second order in two independent variables.
We present a discrete version of the transformation 
for a 6-point difference scheme.
The scheme is appropriate to solving a hyperbolic
type initial-boundary value problem.
We discuss several reductions and specifications of the transformations
as well as construction of other Darboux covariant schemes by means of existing ones.
In particular we introduce a 10-point scheme which can be regarded
as the discretization of self-adjoint hyperbolic equation.
\end{abstract}

Integrable systems, Jonas transformations, Moutard transformations

\section{Introduction}
One can observe increasing role of difference equations over the past few decades.
Primarily efforts were undertaken to discretize differential equations so that not to lose
the properties (e.g. symmetries) that differential equations exhibit. It turned out quickly that difference equations
in many aspects are richer and more fundamental than their continuous counterparts 
(many interesting structures disappeared under a continuum limit) and difference equations started to be
something more than equations mimetic differential equations. In the present work
we encounter the essential differences between discrete and continuous mathematical structures once more.

The aim of this paper is to complete existing theory of Darboux transformations 
(or better to say Darboux--Moutard transformations \cite{Moutard,Darboux,MatSal} 
and Jonas transformations \cite{Jonas})
for differential equations and what more important to show the impact  of the generalization
on the theory of 
Darboux--Moutard transformations
for difference equations.

The main idea of this paper is to start systematic surveys that  can free theory of integrable systems
from their strong dependence of coordinate systems (parametrization of surfaces), desirable by many physicists author spoke to.
It is especially important for the so called difference geometry \cite{Sau} since
due to results of the paper
in the case of difference equations one can compensate  lack of possibility of change of independent
variables $\tilde{x}=f(x,y), \quad \tilde{y}=g(x,y)$.

We recall that classical fundamental transformation by Jonas \cite{Jonas,Eisen} regarded as the most general Darboux transformation
 acts on the conjugate nets in projective space ${\mathbb P}^n$  
so in case when the net is two-dimensional it provide us with Darboux transformation for
two-dimensional linear hyperbolic differential equation in canonical form
(from now on, unless otherwise stated, small letters  denote functions of real independent variables $x$ and $y$
and subscripts foregone 
by comma denote partial differentiation with respect to indicated variables)
\begin{equation} 
\label{cn}
\psi,_{xy}+ w \psi,_x +z \psi,_y+f \psi=0
\end{equation}
and the transformation is nothing but the spatial part of (binary) Darboux--B\"acklund transformation for 
two-component KP hierarchy.
When the net is more than two-dimensional the Jonas fundamental transformation 
provides us with Darboux transformation for the set of compatible equations of the form \mref{cn}
and  serves  as spatial part of binary Darboux--B\"acklund transformation for
multicomponent KP hierarchies  (in other words yields B\"acklund transformation for
n-wave interaction equations called sometimes Darboux equations see e.g. \cite{Kac,Dolo}).

The fundamental transformation has been successfully translated into the discrete
language \cite{DS,MDS,DSM}. The discrete counterpart of conjugate nets are quadrilateral lattices
in ${\mathbb P}^n$ 
governed by system of equations of the type 
(unless otherwise stated in the whole paper capital letters denote functions of two discrete variables  
$m$ and $n$ ($(m,n)\in {\mathbb Z}^2$), 
$\Di$ and $\Dj$ denotes forward difference operators
$\Di \Psi:= \Psi_{m+1,n}-\Psi$ and
$\Dj \Psi:= \Psi_{m,n+1}-\Psi$
while $\Dim$ and $\Djm$ denotes backward difference operators
$\Dim \Psi:= \Psi_{m-1,n}-\Psi$ and
$\Djm \Psi:= \Psi_{m,n-1}-\Psi$, note we identify $\Psi\equiv\Psi_{m,n}$ and in the whole paper we apply this
convention)
\begin{equation} 
\Di \Dj \Psi+ A \Di \Psi + B \Dj \Psi +C \Psi = 0 
\end{equation}
That is why in recent years notion  of integrability of discrete (difference) equations
was often related to the planarity - 4-point schemes were the building blocks of the theory.

Generalization to the so called quad-graphs  (planar graphs), 
still objects with incorporated planarity, appeared only recently \cite{Adler,BS,BS2,ABS}.

In the present paper we show that planarity is not crucial
from the point of view of integrable systems. It is 
remarkable but only example of more general theory with a 6-point difference scheme 
and 7-point self-adjoint difference scheme
as building blocks of the discrete theory of Darboux-Moutard transformations.

\begin{figure}[!ht]
\centering \fboxsep=10mm \fbox{
\begin{picture}(120,100)
\put(62,92){\circle*{5}}
\put(31,46){\circle*{5}} \put(93,46){\circle*{5}}
\put(0,0){\circle*{5}} \put(62,0){\circle*{5}}  \put(122,0){\circle*{5}} 
\put(65,98){$\jjj$}
\put(34,52){$\jj$} \put(96,52){$\ij$}
\put(3,6){$\ioo$} \put(65,6){$\ii$} \put(125,6){$\iii$}
\end{picture}
} 
\caption{The 6-point scheme}
\label{pic6}
\end{figure}
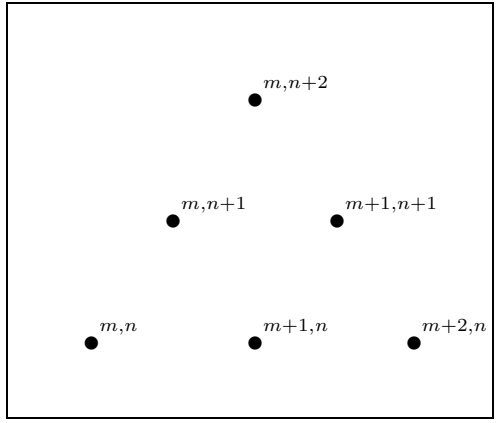

It turns out that the general differential equation
of the second order in two independent variables
\[ \left(a \psi ,_x + c \psi ,_y \right) ,_x+
\left(c \psi ,_x + b \psi ,_y \right) ,_y+
w \psi ,_x + z \psi ,_y -f \psi=0 \]
is covariant under a Darboux transformations  (section \ref{cont})
so conjugate nets are no longer of key importance.
On the discrete level it reflects in the fact that
one can generalize  4-point scheme to a 6-point scheme (see Fig. \ref{pic6})
\[A \Psi\iii  +B \Psi\jjj+2 C \Psi\ij+G \Psi\ii +H \Psi\jj=F \Psi\]
and quadrilateral lattices cease to be the master object of study
 in favour of triangular lattices with the 6-point scheme defined on them (section \ref{disc}).

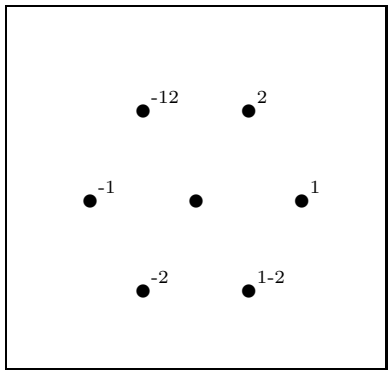
\begin{figure}[!ht]
\centering \fboxsep=10mm \fbox{
\begin{picture}(80,80)
\put(20,69){\circle*{5}} \put(60,69){\circle*{5}}
\put(0,35){\circle*{5}}\put(40,35){\circle*{5}} \put(80,35){\circle*{5}}
\put(20,1){\circle*{5}} \put(60,1){\circle*{5}}
\put(23,72){\scriptsize -12} \put(63,72){\scriptsize 2}
\put(3,38){\scriptsize -1 }\put(43,38){\scriptsize  } \put(83,38){\scriptsize 1 }
\put(23,4){\scriptsize -2 } \put(63,4){\scriptsize 1-2}
\end{picture}
} \caption{7-point scheme}
\label{pic7}
\end{figure}

Moutard transformation for the 7-point self-adjoint scheme (see Fig. \ref{pic7})
\begin{displaymath}
\begin{array}{c}
 {\mathcal A}\ii \, N \ii + {\mathcal A} \, N\iim  +  
 {\mathcal B}\jj \, N \jj + {\mathcal B} \, N\jjm  +\\
 {\mathcal C}\ii N\jijm+{\mathcal C}\jj N\jimj
={\mathcal F} \, N, 
\end{array}
\end{displaymath}
an example of equation given on a star (cross),
has been derived 
in the paper \cite{NSD} (see also subsection \ref{7}) and that is why we concentrate
here  mainly on the 6-point scheme.
However, the present paper is thought to provide a brief overview on the topic of discretizations of 2D second order differential
equations that are covariant under a Darboux transformations
and the reader can find in the closing section \ref{discuss} 
references to articles on integrable aspects of equations given on stars.
At the moment
we only underline that choice of difference scheme restricts sorts of initial-boundary value
problems one can solve  by means of the scheme. So it is important to indicate first on sort of
initial-boundary conditions one would like to solve and then consider only the schemes that allow
to solve the initial-boundary value  problem. We start the paper with description
of a well like initial-boundary value  problem (section \ref{boundary}) we have in mind while 6-point scheme is considered.
In turn the 7-point scheme
is not proper  to solve the initial-boundary value problem 
(as we know from numerical analysis   schemes on stars are suitable
to solve numerically Dirichlet boundary value problem for 
an elliptic differential equation, see e.g. \cite{Hild}) but one can construct a proper
scheme using the 7-point scheme (see subsection \ref{discM}).

In this paper we discuss also how the general Darboux transformation
can be reduced or specified. We take the stand that introduction of novel
terminology (such us specification)
is necessary to discern procedures we deal with.
We begin the discussion of reductions and specifications from the continuous case (section \ref{contR}).
Firstly, we consider the Moutard reduction (subsection \ref{contM})
which is  very classical construction \cite{Moutard}
but to the best of our knowledge in full generality was given only recently
\cite{NSD}. Secondly, we discuss specifications (subsection \ref{contS}) 
and this part is
(stands to reasons)
 new, specification to hitherto  considered "conjugate" case
or its  elliptic counterpart are
just  examples of such procedure.
Thirdly, we discuss two convenient gauge specifications (subsection \ref{contG}) the transformations can be written in. 
Finally we discuss reductions and specifications in the discrete
case.

We start from gauge specifications  (subsection \ref{discG}) 
and specifications (subsection \ref{discS}).  
We are not able to show reduction of the general 
6-point scheme
that leads to transformation of the Moutard type. 
Therefore we recall all results related to the topic first  
(section \ref{discMr}) and then
all we are able to do
is to introduce a 10-point scheme which is appropriate
for solving defined in section \ref{boundary} initial boundary value problem  and can be regarded
as  a discretization of self-adjoint differential equation
(section \ref{discM}).

We would like to stress once more that
although we deal in the paper with linear equations only, the existence of Darboux-Moutard transformations
makes this paper especially important for the theory of nonlinear integrable systems.

\section{Well like  initial-boundary value problem for 6-point scheme}
\label{boundary}
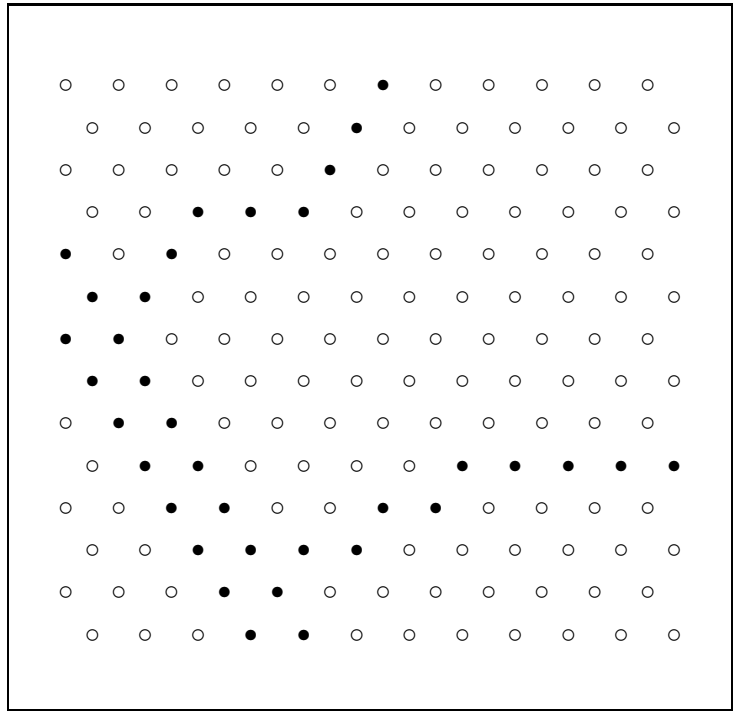
\begin{figure}[!ht]
\centering \fboxsep=10mm \fbox{
\begin{picture}(210,210)
\put(220,0){\circle{4}} \put(200,0){\circle{4}} \put(180,0){\circle{4}}\put(160,0){\circle{4}}
 \put(140,0){\circle{4}} \put(120,0){\circle{4}}\put(100,0){\circle{4}}\put(80,0){\circle*{4}}
  \put(60,0){\circle*{4}} \put(40,0){\circle{4}}\put(20,0){\circle{4}}\put(0,0){\circle{4}}
\put(210,16){\circle{4}} \put(190,16){\circle{4}}\put(170,16){\circle{4}}\put(150,16){\circle{4}}
 \put(130,16){\circle{4}} \put(110,16){\circle{4}}\put(90,16){\circle{4}}\put(70,16){\circle*{4}}
  \put(50,16){\circle*{4}} \put(30,16){\circle{4}}\put(10,16){\circle{4}}\put(-10,16){\circle{4}}
\put(220,32){\circle{4}} \put(200,32){\circle{4}} \put(180,32){\circle{4}}\put(160,32){\circle{4}}
 \put(140,32){\circle{4}} \put(120,32){\circle{4}}\put(100,32){\circle*{4}}\put(80,32){\circle*{4}}
  \put(60,32){\circle*{4}} \put(40,32){\circle*{4}}\put(20,32){\circle{4}}\put(0,32){\circle{4}}
\put(210,48){\circle{4}} \put(190,48){\circle{4}}\put(170,48){\circle{4}}\put(150,48){\circle{4}}
 \put(130,48){\circle*{4}} \put(110,48){\circle*{4}}\put(90,48){\circle{4}}\put(70,48){\circle{4}}
  \put(50,48){\circle*{4}} \put(30,48){\circle*{4}}\put(10,48){\circle{4}}\put(-10,48){\circle{4}} 
\put(220,64){\circle*{4}} \put(200,64){\circle*{4}} \put(180,64){\circle*{4}}\put(160,64){\circle*{4}}
 \put(140,64){\circle*{4}} \put(120,64){\circle{4}}\put(100,64){\circle{4}}\put(80,64){\circle{4}}
  \put(60,64){\circle{4}} \put(40,64){\circle*{4}}\put(20,64){\circle*{4}}\put(0,64){\circle{4}}
\put(210,80){\circle{4}} \put(190,80){\circle{4}}\put(170,80){\circle{4}}\put(150,80){\circle{4}}
 \put(130,80){\circle{4}} \put(110,80){\circle{4}}\put(90,80){\circle{4}}\put(70,80){\circle{4}}
  \put(50,80){\circle{4}} \put(30,80){\circle*{4}}\put(10,80){\circle*{4}}\put(-10,80){\circle{4}} 
\put(220,96){\circle{4}} \put(200,96){\circle{4}} \put(180,96){\circle{4}}\put(160,96){\circle{4}}
 \put(140,96){\circle{4}} \put(120,96){\circle{4}}\put(100,96){\circle{4}}\put(80,96){\circle{4}}
  \put(60,96){\circle{4}} \put(40,96){\circle{4}}\put(20,96){\circle*{4}}\put(0,96){\circle*{4}}
\put(210,112){\circle{4}} \put(190,112){\circle{4}}\put(170,112){\circle{4}}\put(150,112){\circle{4}}
 \put(130,112){\circle{4}} \put(110,112){\circle{4}}\put(90,112){\circle{4}}\put(70,112){\circle{4}}
  \put(50,112){\circle{4}} \put(30,112){\circle{4}}\put(10,112){\circle*{4}}\put(-10,112){\circle*{4}}
\put(220,128){\circle{4}} \put(200,128){\circle{4}} \put(180,128){\circle{4}}\put(160,128){\circle{4}}
 \put(140,128){\circle{4}} \put(120,128){\circle{4}}\put(100,128){\circle{4}}\put(80,128){\circle{4}}
  \put(60,128){\circle{4}} \put(40,128){\circle{4}}\put(20,128){\circle*{4}}\put(0,128){\circle*{4}}
\put(210,144){\circle{4}} \put(190,144){\circle{4}}\put(170,144){\circle{4}}\put(150,144){\circle{4}}
 \put(130,144){\circle{4}} \put(110,144){\circle{4}}\put(90,144){\circle{4}}\put(70,144){\circle{4}}
  \put(50,144){\circle{4}} \put(30,144){\circle*{4}}\put(10,144){\circle{4}}\put(-10,144){\circle*{4}}
\put(220,160){\circle{4}} \put(200,160){\circle{4}} \put(180,160){\circle{4}}\put(160,160){\circle{4}}
 \put(140,160){\circle{4}} \put(120,160){\circle{4}}\put(100,160){\circle{4}}\put(80,160){\circle*{4}}
  \put(60,160){\circle*{4}} \put(40,160){\circle*{4}}\put(20,160){\circle{4}}\put(0,160){\circle{4}}
\put(210,176){\circle{4}} \put(190,176){\circle{4}}\put(170,176){\circle{4}}\put(150,176){\circle{4}}
 \put(130,176){\circle{4}} \put(110,176){\circle{4}}\put(90,176){\circle*{4}}\put(70,176){\circle{4}}
  \put(50,176){\circle{4}} \put(30,176){\circle{4}}\put(10,176){\circle{4}}\put(-10,176){\circle{4}}
\put(220,192){\circle{4}} \put(200,192){\circle{4}} \put(180,192){\circle{4}}\put(160,192){\circle{4}}
 \put(140,192){\circle{4}} \put(120,192){\circle{4}}\put(100,192){\circle*{4}}\put(80,192){\circle{4}}
  \put(60,192){\circle{4}} \put(40,192){\circle{4}}\put(20,192){\circle{4}}\put(0,192){\circle{4}}
\put(210,208){\circle{4}} \put(190,208){\circle{4}}\put(170,208){\circle{4}}\put(150,208){\circle{4}}
 \put(130,208){\circle{4}} \put(110,208){\circle*{4}}\put(90,208){\circle{4}}\put(70,208){\circle{4}}
  \put(50,208){\circle{4}} \put(30,208){\circle{4}}\put(10,208){\circle{4}}\put(-10,208){\circle{4}}
\end{picture}
} \caption{Initial-boundary value problem. The initial values  at points of two neighbouring straight line are given
as well as two boundary conditions (black points).
}
\label{pic8}
\end{figure}

In the present paper we pay special attention  to
difference schemes that allows to solve the following
initial boundary value problem.
We prescribe function $\Psi(m,n)$ in the following
points of the domain (see Fig. \ref{pic8})
\begin{itemize}
\item initial conditions
\[ \{ (m,n) \in {\mathbb T} | \, m+n=0 \vee m+n=1 \} \]
\item boundary conditions
\[ \{ (m,n) \in {\mathbb T} | \, (m=s-p_s \wedge n=p_s) \vee (m=s-p_s \wedge n=p_s), s=2,3,4,...\}\]
\end{itemize}
where ${\mathbb T}$ denotes regular triangular lattice, $p_s$ and $q_s$ are functions 
${\mathbb N}\backslash \{1\} \ni s\mapsto p_s \in {\mathbb Z}$ 
such that $ \forall s \in{\mathbb N}\backslash \{1\} p_s<q_s $

We concentrate in the paper mainly on the schemes that allow to find solution uniquely 
at least in the "upper half-plane" $\{ (m,n) \in {\mathbb T} | \, m+n \ge 0 \}$ of the lattice.

For instance in the case of the 6-point scheme if the following conditions are satisfied 

1) \[ \forall (m,n) \in{\mathbb N} \times{ \mathbb N}, \, \, A_{m,n} \ne 0, B_{m,n} \ne 0, F_{m,n} \ne 0  \]

2) \[ \forall s \in{\mathbb N}\backslash \{1\} \] the matrices 
\begin{displaymath}
\label{codition}
\left[ 
\begin{matrix}
2 C_{s-p_s-2,p_s} & B_{s-p_s-2,p_s}   &\! \! \! \! \! \! \! \! \! 0                 & \! \! \!   \cdots &  & 0\cr
A_{s-p_s-3,p_s+1} &2C_{s-p_s-3,p_s+1} & B_{s-p_s-3,p_s+1} & \! \! \!  0    &  & 0\cr
0                 &                   &                   & \! \! \!       &\ddots& \vdots\cr
\vdots &\ddots&& \! \! \! &&0 \cr
0 & \cdots &\! \! \! \! \! \! \! \! \! 0&\! \! \! \! \! \! \! \! \! \! \! \! \! \! \! \! \! \! A_{s-q_s+1,q_s-1} &2C_{s-q_s+1,q_s-1} & B_{s-q_s+1,q_s-1} \cr
0& &\! \! \! \! \! \!  \cdots &   \! \! \! 0& \! \! \! A_{s-q_s,q_s-2}&2C_{s-q_s,q_s-2} 
\end{matrix}
\right]
\end{displaymath}
have non-vanishing determinant,
then all the values at white points can be found uniquely.
Similar result  can be obtain for the 10-point scheme \mref{10} with the only essential
difference that the solution can be found uniquely only in upper half-plane.

\section{Darboux transformations for 2D second order differential equation}
\label{cont}

It is a basic observation that map $\psi^t \mapsto
\bar{\psi}^t$ given by
\begin{equation}
\label{gct}
\begin{array}{c}
\bar{\psi}^t,_x= \delta \psi^t,_x+\beta  \psi^t,_y \qquad 
\bar{\psi}^t,_y= -\alpha \psi^t,_x-\gamma \psi^t,_y,
\end{array}
\end{equation}
where $\alpha$, $\beta$, $\gamma$ and $\delta$ are 
${\mathcal C}^1$ real functions
(of independent variables $x$ and $y$ with an open, simply connected  subset $\mathcal D$ of ${\mathbb R}^2$ as a domain) such that $\forall (x,y)
\in {\mathcal D},  \,\,\, \gamma \delta -
\alpha \beta \ne 0 $ and $\alpha^2+ \beta^2 + (\gamma + \delta)^2 \ne 0 $, 
is an invertible map between
solution spaces of two differential equations of second
order in two independent variables. Indeed, the compatibility
condition of \mref{gct}, which ensures existence of 
$\bar{\psi}^t$ function, reads
\begin{equation}
\label{etf}
\begin{array}{c}
{\mathcal L}^t \psi^t =0 \\ \\
{\mathcal L}^t := \alpha \partial^2_x+\beta \partial^2_y+
(\gamma + \delta) \partial_x \partial_y +
(\alpha,_{x}+\delta,_{y}) \partial_x+
(\beta,_{y}+\gamma,_{x}) \partial_y.
\end{array}
\end{equation}
Obviously $\bar{\psi}^t$  satisfies equation of the same type but
with bared coefficients
\begin{equation}
\begin{array}{c}
\bar{\alpha}=\frac{\alpha}{\gamma \delta - \alpha\beta}, \quad
\bar{\beta} =\frac{\beta}{ \gamma \delta - \alpha\beta}, \quad
\bar{\gamma}=\frac{\delta}{\gamma \delta - \alpha\beta}, \quad
\bar{\delta}=\frac{\gamma}{\gamma \delta - \alpha\beta}.
\end{array}
\end{equation}
As we shall see every second order equation in two
independent variables
\begin{equation}
\label{f}
\begin{array}{c}
{\mathcal L}^f \psi =0 \\ \\
{\mathcal L}^f :=
a \partial^2_x + b \partial^2_y+ 2 c \partial_x \partial_y+
(a,_x+c,_y+w)
\partial_x+(b,_y+c,_x+z)\partial_y-f 
\end{array}
\end{equation}
can be transformed into the form \mref{etf} through a gauge transformation
\begin{equation}
\label{gt}
{\mathcal L}^f \mapsto {\mathcal L}:= \hat{\phi} {\mathcal L}^f \hat{\theta}
\end{equation}
where $\hat{\phi}$ and $\hat{\theta}$ are operators of multiplying by function
$\phi(x,y)$  and $\theta(x,y)$ respectively, which we are going to 
determine now. We will call operator ${\mathcal L}^t$ 
{\em elementary transformable form}
of the second order differential operator.

Indeed, request that operator ${\mathcal L}$ defined in eq. \mref{gt} is of the
elementary transformable form ${\mathcal L}^t$ \mref{etf} gives
\begin{equation}
\label{c1}
\alpha= \phi \theta a \qquad \beta=\phi \theta b  \qquad 
\gamma + \delta = 2 \phi \theta  c  
\end{equation}
\begin{equation}
\label{c2}
{\mathcal L}^f \theta=0
\end{equation}
\begin{equation}
\label{c3}
\begin{array}{lll}
\alpha ,_x+\delta ,_y & = &
[ (a \theta),_x + (c \, \theta),_y + a \theta ,_x +c \, \theta ,_y +w \, \theta ] \phi,
\\ 
\beta ,_y+\gamma ,_x & = &
[(c \, \theta),_x + (b \, \theta),_y + c \, \theta ,_x +b \, \theta ,_y +z \, \theta ] \phi.
\end{array}
\end{equation}
On introducing auxiliary function $p$
\begin{equation}
\label{c4}
p:= \frac{1}{2\phi \theta} ( \delta - \gamma) 
\end{equation}
and treating  eqs \mref{c1}, \mref{c4} as the definition of functions
$\alpha$, $\beta$, $\gamma$ and $\delta$ and eliminating these functions
from equations \mref{c3} one can rewrite equations \mref{c3} in the form
\begin{equation}
\label{p}
\begin{array}{llll}
(\theta \phi p),_y&=&&
\phi^2 [a (\frac{\theta}{\phi}),_x + c (\frac{\theta}{\phi}),_y +w \frac{\theta}{\phi}]
\\
(\theta \phi p),_x&=&-&
\phi^2 [c (\frac{\theta}{\phi}),_x + b (\frac{\theta}{\phi}),_y +z \frac{\theta}{\phi}]
\end{array}
\end{equation}
Function $p$ exists provided that
\begin{equation}
\label{c6}
\phi {\mathcal L}^f \theta - \theta ({\mathcal L}^f)^{\dagger} \phi=0
\end{equation}
Where $({\mathcal L}^f)^{\dagger}$ denotes the operator formally adjoint to the operator ${\mathcal L}^f$
\begin{equation}
\label{fad}
\begin{array}{c}
({\mathcal L}^f)^{\dagger} :=
\partial_x(a \partial_x + c  \partial_y -w)+\partial_y(b \partial_y + c  \partial_x -z)-f 
\end{array}
\end{equation}
Taking into account \mref{c2} and \mref{c6} we come to the theorem

\begin{thm}
\label{thc}
Gauge transformation \mref{gt} makes from arbitrary 2D second order operator 
${\mathcal L}^f$ an operator in elementary transformable form   
{\em iff}
\begin{equation}
\label{concl}
{\mathcal L}^f \theta =0  \qquad \hbox{and} \qquad  ({\mathcal L}^f)^{\dagger} \phi=0
\end{equation}
\end{thm}
It turns out that just presented considerations lead to Darboux transformations for
the 2D second order operator ${\mathcal L}^f$.
Namely we have conclusion

\begin{conc}[Darboux transformations]
We assume that $\theta$ and $\phi$ are ${\mathcal C}^2$ class functions satisfying conditions \mref{concl},
function $p$ is given by formulae \mref{p},  $r$ and $s$ are arbitrary (of class ${\mathcal C}^2$) functions, and function $d$  given by $d:=(p^2-c^2+ab) \phi \, \theta$
obeys condition $\forall (x,y) \in {\mathcal D}; d \ne 0$.
Then the map $\psi \mapsto \bar{\psi}$ given by

\begin{equation}
\label{matrc}
\left[ 
\begin{matrix}
(s \bar{\psi}),_x \cr (s \bar{\psi}),_y
\end{matrix}
\right] = \phi \theta
\left[ 
\begin{matrix}
p+c&b\cr
-a&p-c
\end{matrix}
\right]
\left[ 
\begin{matrix}
\left(\frac{\psi}{\theta} \right),_x \cr \left(\frac{\psi}{\theta} \right),_y
\end{matrix}
\right]
\end{equation}
is the map from solution space of equation \mref{f} to the solution space of
the equation of the same form but with the new coefficients

\begin{equation}
\label{LB}
\begin{array}{c}
\bar{{\mathcal L}} \bar{\psi} =0 \\
\bar{{\mathcal L}}:=\bar{a} \partial_x^2 + \bar{b} \partial_y^2 +2 \bar{c}\partial_x\partial_y
+(\bar{a},_x+\bar{c},_y+\bar{w})\partial_x +(\bar{c},_x+\bar{b},_y+\bar{z})\partial_y - \bar{f}
\end{array}
\end{equation}
where the coefficients of (\ref{LB}) are related to coefficients of (\ref{f}) by
\begin{equation}
\label{bar}
\begin{array}{c}
\bar{a}=\frac{a s r}{d}, \qquad \bar{b}=\frac{b s r}{d},\qquad \bar{c}=\frac{csr}{d},\\
\\
\bar{w}=[\frac{a}{d} (\frac{s}{r}),_x+\frac{c}{d} (\frac{s}{r}),_y
-(\frac{p}{d}),_y \frac{s}{r}] r^2, \quad
\bar{z}=[\frac{b}{d} (\frac{s}{r}),_y+\frac{c}{d} (\frac{s}{r}),_x
+(\frac{p}{d}),_x \frac{s}{r}] r^2. \\
\\
\bar{f}=\{
-[\frac{a}{d} s_{x}+\frac{c+p}{d} s_{y}]_x
-[\frac{b}{d} s_{y}+\frac{c-p}{d} s_{x}]_y
\} r
\end{array}
\end{equation}
\end{conc}

\section{ 6--point scheme and its Darboux transformations}
\label{disc}

One can repeat considerations from the previous section in the discrete case.
Indeed, starting from pair of equations
\begin{equation}
\label{6e}
\begin{array}{l}
\Di \bar{\Psi}^t = \delta \Di \Psi^t+\beta  \Dj \Psi^t \qquad 
\Dj \bar{\Psi}^t = -\alpha \Di \Psi^t-\gamma \Dj \Psi^t
\end{array}
\end{equation}
(where the functions $\alpha$, $\beta$, $\gamma$ and $\delta$ are functions of discrete variables
$m$ and $n$)
and writing down their compatibility condition
\begin{equation}
\label{deif}
\begin{array}{l}
\alpha \ii \Psi^t \iii   + \beta \jj \Psi^t \jjj + 
(\gamma \ii+ \delta \jj) \Psi^t\ij- \\ 
(\alpha \ii+\alpha  + \gamma \ii+ \delta) \Psi^t \ii 
-(\beta \jj + \beta  +\gamma +\delta \jj) \Psi^t \jj+ \\
(\alpha + \beta + \gamma +\delta ) \Psi^t =0
\end{array}
\end{equation}
which is a 6-point scheme.
One can ask if it is possible to transform the general 6-point scheme of this type
\begin{equation}
\begin{array}{l}
\label{6}
A \Psi\iii +B \Psi\jjj+2 C \Psi\ij+G \Psi\ii +H \Psi\jj=F \Psi
\end{array}
\end{equation}
\[ L^F \Psi = 0 \]
into the form  \mref{deif} via a gauge transformation
\begin{equation}
\label{dgt}
L^F \mapsto L:= \hat{\Phi} L^F \hat{\Theta}
\end{equation}
only? The answer is positive and we will be calling equation of type \mref{deif} a 6-point scheme in {\em elementary transformable form}.
\begin{thm}
Gauge transformation \mref{dgt} makes from the 6-point scheme \mref{6}
$L^f$ an operator in elementary transformable form $L^t$  {\em iff}
the function $\Theta$ satisfies eq. \mref{6} 
\begin{equation}
\begin{array}{l}
\label{6t}
A \Theta\iii  +B \Theta\jjj+2 C \Theta\ij+G \Theta\ii +H \Theta\jj=F \Theta
\end{array}
\end{equation}
while function
$\Phi$ is  a solution of the equation formally adjoint to eq. \mref{6} 
\begin{equation}
\begin{array}{l}
\label{6ad}
A\iimim \Phi\iimim +B\jjmjm \Phi\jjmjm+2 C\jimjm \Phi\jimjm
+\\ G\iim  \Phi\iim  +H\jjm  \Phi\jjm =F \Phi
\end{array}
\end{equation}
Then the functions $\alpha$, $\beta$, $\gamma$ and $\delta$ in eq. (\ref{deif}) are given by
\begin{equation}
\begin{array}{l}
\alpha = A\iim  \Phi\iim  \Theta\ii\\
\beta  = B\jjm  \Phi\jjm  \Theta\jj\\
\gamma = (C\iim -P\iim )\Phi\iim \Theta\jj\\
\delta = (C\jjm +P\jjm )\Phi\jjm \Theta\ii
\end{array}
\end{equation}
where $P$ is an auxiliary function defined by
\begin{equation}
\label{P}
\begin{array}{l}
\Dim (\Phi\Theta \ij P) = 
-(B\jjm  \Phi\jjm  \Theta\jj+B \Phi \Theta\jjj+\\
\hfill{C\iim \Phi\iim \Theta\jj+C \Phi \Theta\ij+H \Phi \Theta\jj)} \\
\Djm (\Phi\Theta \ij P) =
A\iim  \Phi\iim  \Theta\ii+A \Phi \Theta\iii +\\
\hfill{C\jjm \Phi\jjm \Theta\ii+C \Phi \Theta\ij+G \Phi \Theta\ii}
\end{array}
\end{equation}
\end{thm}
We have the conclusion
\begin{conc}[Darboux transformations for the 6-point scheme]
~ \newline
The map $\Psi \mapsto \bar{\Psi}$ given by
\begin{equation}
\label{DT6}
\begin{array}{l}
\left[ 
\begin{matrix}
\Di (S \bar{\Psi}) \cr \Dj (S \bar{\Psi})
\end{matrix}
\right] = \\
\! \! \! \! \! \! \! \! \! \! \left[ 
\begin{matrix}
\! (C\jjm \!+\! P\jjm )\Phi\jjm \Theta\ii& B\jjm  \Phi\jjm  \Theta\jj\cr
\! \! \! \! \! \! \! \! \! \! \! \! \! \! \! \! \! \! \!
\! -A\iim \Phi\iim  \Theta\ii&\! \! \! \! \! \! \! \!\! \! \! \!\! \! \! \! \! \! \! \!(P\iim \!-\! C\iim )\Phi\iim \Theta\jj) \!
\end{matrix}
\right] \! \! \! 
\left[ 
\begin{matrix}
\Di \left(\frac{\Psi}{\Theta} \right)\cr  \Dj \left(\frac{\Psi}{\Theta} \right)
\end{matrix}
\right]
\end{array}
\end{equation}
where function $\Theta$ satisfies eq. \mref{6} 
while function
$\Phi$ is  a solution of the  eq. \mref{6ad} and $P$ is defined via equations
\mref{P}, is the map from solution space of the equation \mref{6} to the solution space of the
equation of the same form with novel (bared) coefficients related to the old ones via
\begin{equation}
\label{dcoef}
\begin{array}{l}
\bar{A}=\frac{R S\iii  \Phi \Theta\iii }{D\ii} A,
\qquad
\bar{B}=\frac{R S\jjj \Phi \Theta\jjj }{D\jj} B,
\qquad
\bar{F}=\frac{R S \Phi \Theta }{D} F,
\\ 
\\
\frac{2 \bar{C}}{R S\ij} =
\frac{\Theta\iii \Phi\jijm(C+P)\jijm}{D\ii}+\frac{\Theta\jjj\Phi\jimj(C-P)\jimj}{D\jj}, 
\\
\\
\frac{\bar{G}}{R S\ii}=-\frac{\Theta\iii \Phi\jijm(C+P)\jijm+\Theta\iii \Phi A}{D\ii}-\\ \\
\hfill{ + \frac{\Theta\jj\Phi\iim (C-P)\iim +\Theta\ii\Phi\iim A\iim }{D}},
\\
\\
\frac{\bar{H}}{R S\jj}=-\frac{\Theta\jjj\Phi\jimj(C-P)\jimj+\Theta\jjj\Phi B}{D\jj}-\\ \\
\hfill{+\frac{\Theta\ii\Phi\jjm (C+P)\jjm +\Theta\jj\Phi\jjm B\jjm }{D}},
\end{array}
\end{equation}
where $R$ and  $S$
are arbitrary non-vanishing functions, while $D$ is a function given by
\newline
{\small$D=[(P\iim \!-\! C\iim )(P\jjm \!+\! C\jjm ) \!+\! A\iim  B\jjm ] \Theta\ii \Theta\jj \Phi\iim  \Phi\jjm $}
and is assumed not to vanish on the whole lattice.
\end{conc}

\section{Gauge specifications, specifications and reductions}
\label{contR}
One can reduce (or gauge or specify) the Darboux transformation in such a way that
it maps between solution space of restricted class of equations.
Assume first that matrix coefficients in eq. \mref{matrc} and coefficients of its inverse 
obey the same linear constraint
\begin{equation}
\begin{array}{l}
\label{aij}
a^{11} (p+c) +a^{12} b-a^{21} a+ a^{22} (p-c)=0 \\
a^{11} (p-c) -a^{12} b+a^{21} a+ a^{22} (p+c)=0
\end{array}
\end{equation}
where $a^{ij}$ are given functions of $x$ and $y$.
From \mref{aij} we infer
\[(a^{11}+a^{22})p=0\] and will discuss two cases $p=0$ and 
$a^{11}+a^{22}=0$ separately.

\subsection{Moutard reduction ($p=0$), reductions}
\label{contM}
We have $p=0$ and $(a^{11}-a^{22}) c +a^{12} b-a^{21} a =0$ to be satisfied. The later
constraint can be satisfied if one take $a^{11}=a^{22}$, $a^{12}=0$ and $a^{21}=0$.
To satisfy the equations \mref{p} in the presence of condition $p=0$ it is enough
to put $\phi=\theta$ and $w=0=z$ (i.e. demand that operator is formally self-adjoint).
Moreover the functions $r$ and $s$ are no longer arbitrary, they must obey constraint
$\frac{r}{s}=const$.
As a result we obtain a transformation for formally self-adjoint equations which
are usually refereed to (in the case $a=0=b$, $c=\frac{1}{2}$, $s=r=\frac{1}{2}\theta$) as Moutard transformation.

The procedure that impose the constraints on transformation's data $\phi$ and $\theta$
($\theta=\phi$ in the Moutard case) we call reduction of the transformation. 

\subsection{Specifications $a^{11}+a^{22}=0$}
\label{contS}
The reduction is  not the only procedure we have to our disposal. If we take $a^{11}=-a^{22}$
we have the constraint $2 a^{11} c +a^{12} b-a^{21} a =0$ to be satisfied.
Two examples are

a) $a=0=b$, $a^{11}=0$ and $c=1$ which is nothing but specification to "conjugate" case

b) $c=0$,  $a^{12}=0=a^{21}$ (with option for deeper specification $a=\pm b$)

In those cases we only specify (specialize) the operator not affecting the transformation data.

\subsection{Gauge specifications, affine form, elementary transformable form}
\label{contG}
The idea not to consider the operator itself  
but the equivalence classes with respect to the gauge
goes back to Laplace and Darboux papers \cite{Laplace}. 
One can then develop theory in gauge independent language or
choose a gauge appropriate to ones needs. We concentrate on the later procedure 
We take two arbitrary functions $\theta^0$, $\phi^0$ of ${\mathcal C}^2$ class. 
Then operator
\[L^g:= \hat{\phi^0} L^f \hat{\theta^0}\]
has coefficients
\[(a^g,b^g,c^g,f^g)=\theta^0 \phi^0(a,b,c,L^f\theta^0) \]

\[w^g=\theta^0 \phi^0 w 
- (\theta^0)^2 (\frac{\phi^0}{\theta^0}),_x a
- (\theta^0)^2 (\frac{\phi^0}{\theta^0}),_y c\]

\[z^g=\theta^0 \phi^0 z 
- (\theta_0)^2 (\frac{\phi^0}{\theta^0}),_y b
- (\theta_0)^2 (\frac{\phi^0}{\theta^0}),_x c\]
and operator adjoint to $L^g$ is
\[(L^g)^\dagger:= \hat{\theta^0} L^f \hat{\phi^0}\]
If in addition we define 
\begin{equation}
\begin{array}{ll}
\psi^g=\frac{\psi}{\theta_0} \quad \phi^g=\frac{\phi}{\phi_0} \quad
\theta^g=\frac{\theta}{\theta_0} 
\end{array}
\end{equation}
then the form of the transformation remains unaltered
\begin{equation}
\label{pAt}
\begin{array}{llll}
(\theta^g \phi^g p^g),_y&=&&
(\phi^g)^2 [a^g (\frac{\theta^g}{\phi^g}),_x + c^g (\frac{\theta^g}{\phi^g}),_y -w^g \frac{\theta^g}{\phi^g}]
\\
(\theta^g \phi^g p^g),_x&=&-&
(\phi^g)^2 [c^g (\frac{\theta^g}{\phi^g}),_x + b^g (\frac{\theta^g}{\phi^g}),_y -z^g \frac{\theta^g}{\phi^g}]
\end{array}
\end{equation}
\begin{equation}
\left[ 
  \begin{matrix}
  (s^g \bar{\psi}^g),_x \cr (s^g \bar{\psi}^g),_y
  \end{matrix}
\right] 
= \phi^g \theta^g
\left[ 
  \begin{matrix}
   p^g+c^g&b^g\cr
   -a^g&p^g-c^g
  \end{matrix}
\right]
\left[ 
  \begin{matrix}
  \left(\frac{\psi^g}{\theta^g} \right),_x \cr \left(\frac{\psi^g}{\theta^g} \right),_y
  \end{matrix}
\right]
\end{equation}
We indicate on two convenient gauges:

1) Elementary transformable gauge 

\[L^f \theta^0 =0 \qquad (L^g)^\dagger \phi^0=0\]
 This gauge specification of the transformation
to an elementary transformable form
i.e., as we know from theorem \ref{thc}, conditions
\[ f^g=0 \qquad w^g,_x+z^g,_y=0\]
hold. Moreover
the operator $(L^g)^\dagger$ is in elementary transformable form as well.
The functions $s^g$ and $r^g$ are no longer arbitrary. To assure the transformed (bared) equation be in
the elementary transformable form it is enough to put the functions  $s^g$ and $r^g$ to be constant.

2) Affine  gauge is 
\[L^f \theta^0 =0 \]
so only
\[ f^A=0 \]
holds.
Now it is enough to put  $s^A$ to be a constant to obtain bared equation in affine form.
Further convenient gauge specification is possible by putting
\[\phi^0=\theta^0\]
Then the operator $L^g$ has the coefficients
\[(a^A,b^A,c^A,w^A,z^A,f^A)=(\theta_0)^2 (a,b,c,w,z,0) \]

\section{Specifications, discrete case}

\subsection{Gauge specifications}
\label{discG}
In the discrete one can specify the gauge as well. 
Namely, we take two arbitrary functions $\Theta^0$ $\Phi^0$ 
Then operator
\[L^g:= \hat{\Phi^0} L^f \hat{\Theta^0}\]
has coefficients
\begin{displaymath}
\begin{array}{l}
(A^g,B^g,C^g,G^g,H^g,F^g)=
\\
\Phi^0 (A \Theta^0\iii ,B \Theta^0\jjj ,C \Theta^0\ij ,G \Theta^0\ii, H \Theta^0\jj,F \Theta^0 ) 
\end{array}
\end{displaymath}
The two examples of convenient gauges are

1) Affine gauge which we obtain
demanding that $\Theta^0$ satisfies the equation
\begin{equation}
\label{gd1}
L^f \Theta^0=0
\end{equation}
Then coefficients of operator $L^g$ obey constraint
\[A^g+B^g+2C^g+G^g+H^g-F^g=0\]
If one puts $S=const$ then the constrain is preserved under the Darboux transformation.

2)
We demand that $\Phi^0$ satisfies the equation
\begin{equation}
\label{gd2}
(L^f) \dagger \Phi^0=0
\end{equation}
Then coefficients of operator $L^g$ obey constraint
\[A^g\jimj+B^g\jijm+2C^g+G^g\jj+H^g\ii-F^g\ij=0\]
If one puts $R=const$ then the constrain is preserved under the Darboux transformation.

If we apply the conditions \mref{gd1} and \mref{gd2} together
and demand $R=0=S$ then we obtain transformation between
elementary invertible forms of the 6-point scheme.

\subsection{Specifications, quadrilateral lattices, 3-point scheme}
\label{discS}
A glance at transformation laws of the coefficients of the 6-point scheme \mref{dcoef}
provide us with conclusions

A)  Both constraint $A=0$, $B=0$ and $F=0$
are preserved under the Darboux transformation \mref{DT6}.
Constraint $A=0=B$ (or alternatively  $A=0=F$ or $B=0=F$)
is specification to the celebrated 4-point scheme i.e. to quadrilateral lattice case
subject of study of many papers (we confine ourselves to citing articles where Darboux transformations
are considered)
\begin{itemize}
\item Quadrilateral Lattices (Jonas fundamental transformations) \cite{MDS,DSM,LM1,W,M}
\item Circular Lattices and Quadratic Reductions (Ribaucour type transformations) \cite{CDS,KonSch,LM2,Dq}
\item Moutard type transformations \cite{Nimmo,NieL}
\item Symmetric (Goursat) type transformations \cite{SS,DSSYM}
\end{itemize}
Initial boundary value problem suitable for this scheme
 is no longer of the type mentioned
in section \ref{boundary}.

B)  Constraint $C=0$ is not preserved under the Darboux transformation \mref{DT6}.
So we have not got discretization of specification  b) from the section \ref{contS}.

C) 
If we impose $F=0$  together with $A=0=B$ we obtain
Darboux transformation for a 3-point scheme which corresponds to the continuous degenerated case 
$a^2+b^2+c^2=0$.

\section{Discrete Moutard case}
\label{discMr}
The question arises can one find difference scheme which is appropriate
to solve initial boundary conditions  we presented in section \ref{boundary} 
and can be regarded as a discretization
of Moutard reduction?
We were not able to find
such a reduction of 6-point scheme and we are not able
to give a satisfactory non-existence theorem of such a 6-point scheme.
Instead we construct the Darboux covariant 10-point scheme that can be regarded
as a discrete Moutard equation and is suitable to solve initial boundary condition
we described in section \ref{boundary}.

In this section we firstly  recall (for completeness of the paper)
known results i.e. the discrete Moutard equation and
its adjoint which were known only for the quadrilateral specification, so far.
Secondly we recall self-adjoint 7-point scheme which is not
proper to solve mentioned in the section \ref{boundary} initial-boundary value problem.
Finally we introduce a discretization of general Moutard reduction
which is not just a reduction of 6-point scheme but arise from such reformulation
of 7-point scheme that allows for solving the initial-boundary value problem.

\subsection{Discrete Moutard equations}
\label{nim}
On putting  $A=0=B$ (quadrilateral specification),
$2C=-F$  and $G=H=:M F$
the equations \mref{6}, \mref{6t} and \mref{6ad} take respectively form
\begin{equation}
\begin{array}{l}
\label{6M}
F (\Psi \ij +\Psi)= G (\Psi\ii + \Psi\jj)      
\end{array}
\end{equation}

\begin{equation}
\begin{array}{l}
\label{6tM}
F(\Theta\ij+\Theta)= G (\Theta\ii + \Theta\jj) 
\end{array}
\end{equation}

\begin{equation}
\begin{array}{l}
\label{6adM}
F\jimjm \Phi\jimjm+F \Phi= G\iim   \Phi\iim  + G\jjm   \Phi \jjm  
\end{array}
\end{equation}
The crucial observation is:
if the function $\Theta$ satisfies equation \mref{6tM} then the function
$\Phi$ given by
\begin{equation}
\label{PT}
\begin{array}{l}
\Phi:=\frac{1}{F} (\Theta\ii+ \Theta\jj)
\end{array}
\end{equation}
satisfies equation \mref{6adM} \cite{NieL}.
If we put
\[2 P=\frac{\Theta\ii- \Theta\jj}{\Phi}\]
then equations \mref{P} will be automatically satisfied. If in addition we put $S=\Theta$ then
Darboux transformation \mref{DT6} takes form 
\begin{equation}
\begin{array}{l}
\Dj \Theta \bar{\Psi} = \Theta \Theta\jj   \Dj \frac{\Psi}{\Theta}\\
\Di \Theta \bar{\Psi} =- \Theta \Theta\ii   \Di \frac{\Psi}{\Theta}
\end{array}
\end{equation}
which is the discrete Moutard transformation given by Nimmo and Schief \cite{Nimmo}.
Function $\bar{\Psi}$ satisfies the equation
\begin{equation}
\begin{array}{l}
\label{6Mb}
\bar{\Psi}\ij+\bar{\Psi}= \bar{M} (\bar{\Psi}\ii + \bar{\Psi}\jj) 
\end{array}
\end{equation}
where
\begin{equation}
\begin{array}{l}
\label{6tMb}
\bar{M}= \frac{\frac{1}{\Theta\ii} + \frac{1}{\Theta\jj}}{\frac{1}{\Theta\ij}+\frac{1}{\Theta}}
\end{array}
\end{equation}
Equations \mref{6tM} and \mref{6adM} are not the same so we cannot say that we have derived self-adjoint reduction.
That is why we'd prefer to call the reduction Moutard reduction rather than self-adjoint reduction.

\subsection{7-point self-adjoint scheme and its 5-point specification}
\label{7}
To made the paper self-contained we derive in different manner
the results contained in  \cite{NSD}.
In the discrete case one can write
\begin{equation}
\label{7e}
\begin{array}{l}
\Di \bar{N}^t = \delta \Dim N^t+\beta  \Djm N^t \qquad 
\Dj \bar{N}^t = -\alpha \Dim N^t-\gamma \Djm N^t
\end{array}
\end{equation}
so instead of forward difference operators on the right side of equations \mref{6e} we have introduced
backward difference operators
in \mref{7e}. The compatibility condition provide us with an elementary transformable 7-point scheme
\[\Dj (\delta \Dim N^t+\beta  \Djm N^t)+ \Di (\alpha \Dim N^t+\gamma \Djm N^t)=0\]
and function $\bar{N}$ satisfies an elementary transformable 7-point scheme as well.
The question arises can one transform the general 7-point scheme 
\begin{displaymath}
\begin{array}{l}
{\mathcal A}\ii\Psi\ii+{\mathcal G} \Psi\iim +
   {\mathcal B}\jj\Psi\jj+{\mathcal H} \Psi\jjm + \\
  {\mathcal C}\ii \Psi\jijm+ {\mathcal D}\jj \Psi\jimj  ={\mathcal F} \Psi
\end{array}
\end{displaymath}
to an elementary transformable 7-point scheme
by a gauge transformation only?
The answer is negative this time
because apart from equations 
\begin{equation}
\begin{array}{l}
\gamma= {\mathcal G}\iim  \varPhi\iim  \varTheta\jjm , \quad  \delta= {\mathcal H}\jjm  \varPhi\jjm  \varTheta\iim , \\
\alpha= -{\mathcal A} \varPhi\iim  \varTheta- {\mathcal G}\iim  \varPhi\iim  \varTheta\jjm \\ 
\beta= -{\mathcal B} \varPhi\jjm  \varTheta-{\mathcal H}\jjm  \varPhi\jjm  \varTheta\iim 
\end{array}
\end{equation}
one has to satisfy three equations
\begin{displaymath}
\begin{array}{l}
 {\mathcal G} \varPhi \varTheta\iim  +{\mathcal D} \varPhi\jjm  \varTheta\iim 
={\mathcal A} \varPhi\iim  \varTheta +{\mathcal C} \varPhi\iim  \varTheta\jjm \\
 {\mathcal H} \varPhi \varTheta\jjm  +{\mathcal C} \varPhi\iim  \varTheta\jjm 
={\mathcal B} \varPhi\jjm  \varTheta +{\mathcal D} \varPhi\jjm  \varTheta\iim \\
({\mathcal A}\ii \varTheta\ii\! +\! {\mathcal B}\jj \varTheta\jj\! +\! {\mathcal C}\ii \varTheta\jijm\! +\! {\mathcal D}\jj \varTheta\jimj)\varPhi   + \\  \varTheta ({\mathcal A} \varPhi\iim + {\mathcal B}\varPhi\jjm )=
{\mathcal F} \varTheta\varPhi
\end{array}
\end{displaymath}
Happily enough the first two equations of 
can be satisfied by putting 
\[\varPhi=\varTheta , \quad {\mathcal G}={\mathcal A}, \quad {\mathcal H}={\mathcal B}, \quad {\mathcal C}={\mathcal D}\]
while the third one takes form

\begin{equation}
\begin{array}{l}
 {\mathcal A}\ii\varTheta\ii+{\mathcal A} \varTheta\iim +
   {\mathcal B}\jj\varTheta\jj+{\mathcal B} \varTheta\jjm 
   + \\ {\mathcal C}\ii \varTheta\jijm+{\mathcal C}\jj \varTheta\jimj ={\mathcal F} \varTheta
\end{array}
\end{equation}
As a result we obtain Darboux transformation for 7-point self-adjoint scheme (c.f. \cite{NSD})
\begin{thm}
Map $N \mapsto \bar{N}$ given by
\begin{equation}
\begin{array}{l}
\!\!\!\!\!   \left[ 
\begin{matrix}
\Di (\bar{N}  \varTheta) \cr \Dj ( \bar{N} \varTheta) 
\end{matrix}
\right] \!\!=\! \!
\left[ 
\begin{matrix}
 \!\!\!\!\!\!{\mathcal C} \varTheta\iim  \varTheta\jjm &\!\!\!\!\!\!
-\varTheta\jjm  ({\mathcal B} \varTheta  \! +\!
{\mathcal C} \varTheta\iim  ) \cr
\varTheta\iim  \left( {\mathcal A} \varTheta  \!+\! {\mathcal C}
 \varTheta\jjm    \right) &\!\!\!\!
- {\mathcal C} \varTheta\iim  \varTheta\jjm 
\end{matrix}
\right]
\!\!\!\left[ 
\begin{matrix}
\Dim \frac{N}{\varTheta}
\cr
\Djm \frac{N}{\varTheta} 
 \end{matrix}
\right] \hfill{}
\end{array}
\end{equation}
is a map from solution space of equation
\begin{equation}
\begin{array}{l}
\label{b1}
 {\mathcal A}\ii \, N\ii + {\mathcal A} \, N\iim  +  
 {\mathcal B}\jj \, N\jj + {\mathcal B} \, N\jjm  +\\
 {\mathcal C}\ii N\jijm+{\mathcal C}\jj N\jimj
={\mathcal F} \, N, 
\end{array}
\end{equation}
to solution space of equation
\begin{equation}
\begin{array}{l}
\label{bb}
 \bar{{\mathcal A}}\ii \,  \bar{N}\ii + \bar{{\mathcal A}} \, \bar{N}\iim  +  \bar{{\mathcal B}}\jj \, \bar{N}\jj+
 \bar{{\mathcal B}} \, 
\bar{N} \jjm  +\\
\bar{{\mathcal C}}\ii \bar{N}\jijm+\bar{{\mathcal C}}\jj \bar{N}\jimj
=\bar{{\mathcal F}} \, \bar{N}, 
\end{array}
\end{equation}
where $\varTheta$ is a solution of equation \mref{b1} and the new (bared) fields are related to the old ones  by
\begin{equation}
\begin{array}{l}
\bar{{\mathcal A}}\ii=\frac{\varTheta \varTheta\ii {\mathcal A}}{\varTheta\jjm   {\mathcal P}},  
\qquad
\bar{{\mathcal B}}\jj=\frac{\varTheta \varTheta\jj {\mathcal B}}{\varTheta\iim   {\mathcal P}},
\\
\\
\bar{{\mathcal C}}=\frac{{\mathcal C}\jimjm \varTheta\iim  \varTheta\jjm  }
{\varTheta\jimjm {\mathcal P}\jimjm}, 
\\
\\
\bar{{F}}=\varTheta (\bar{{\mathcal A}}\ii \frac{1}{\varTheta\ii}+   
\bar{{\mathcal A}} \frac{1}{\varTheta\iim }+  
\bar{{\mathcal B}}\jj \frac{1}{\varTheta\jj} + \bar{{\mathcal B}} \frac{1}{\varTheta\jjm }+
\\
\\
\bar{{\mathcal C}}\ii \frac{1}{\varTheta\jijm}+\bar{{\mathcal C}}\jj \frac{1}{\varTheta\jimj})
\end{array}
\end{equation}
and where
${\mathcal P}:= \varTheta {\mathcal A}{\mathcal B} +
\varTheta\iim  {\mathcal C} {\mathcal A}+\varTheta\jjm  {\mathcal C}  {\mathcal B}$. 
\end{thm}

Clearly the scheme above admits specification ${\mathcal C}=0$ 
(alternatively one can put ${\mathcal A}=0$ or ${\mathcal B}=0$) and as result we obtain specification
to 5-point self-adjoint scheme (c.f. \cite{NSD}).

\subsection{Discrete Moutard type transformation, a 10-point scheme}
\label{discM}

The idea is that in the Darboux transformation for the 7-point scheme we considered in the previous
subsection  one can replace
field $N$ with $\Psi\ii+\Psi\jj$ without affecting leading terms of continuum limit.
As a result we get
   the 10-point scheme
\begin{equation}
\begin{array}{l}
\label{10}
({\mathcal A}+{\mathcal B}) \Psi + ({\mathcal A}\ii+{\mathcal B}\jj) \Psi\ij+\\
{\mathcal A}\Psi\jimj+{\mathcal B}\Psi\jijm+{\mathcal A}\ii\Psi\iii +{\mathcal B}\jj\Psi\jjj+\\
{\mathcal C}\ii\Psi \iiijm +{\mathcal C}\jj\Psi \iimjj +\\
({\mathcal C}\ii-{F})\Psi\ii+({\mathcal C}\jj-F)\Psi\jj=0
\end{array}
\end{equation}

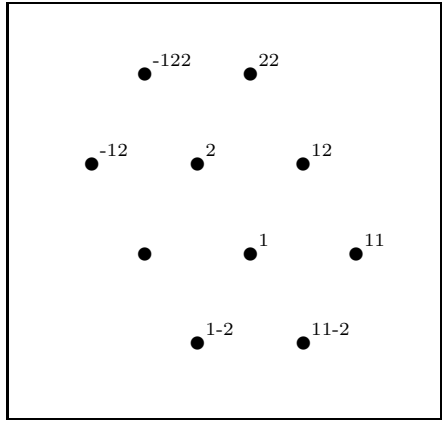
\begin{figure}[!ht]
\centering \fboxsep=10mm \fbox{
\begin{picture}(100,100)
\put(20,102){\circle*{5}} \put(60,102){\circle*{5}}
\put(0,68){\circle*{5}}\put(40,68){\circle*{5}} \put(80,68){\circle*{5}}
\put(20,34){\circle*{5}}\put(60,34){\circle*{5}} \put(100,34){\circle*{5}}
\put(40,0){\circle*{5}} \put(80,0){\circle*{5}}
\put(23,105){\scriptsize -122} \put(63,105){\scriptsize 22}
\put(3,71){\scriptsize -12 }\put(43,71){\scriptsize 2 } \put(83,71){\scriptsize 12 }
\put(23,37){\scriptsize }\put(63,37){\scriptsize 1 } \put(103,37){\scriptsize 11 }
\put(43,3){\scriptsize 1-2 } \put(83,3){\scriptsize 11-2}
\end{picture}
} \caption{10-point scheme}
\label{pic10}
\end{figure}
which is no longer self-adjoint but if the function $\varTheta$ satisfies the equation \mref{10}
\begin{equation}
\begin{array}{l}
({\mathcal A}+{\mathcal B}) \Theta + ({\mathcal A}\ii+{\mathcal B}\jj) \Theta\ij+\\
{\mathcal A}\Theta\jimj+{\mathcal B}\Theta\jijm+{\mathcal A}\ii\Theta\iii +{\mathcal B}\jj\Theta\jjj+ 
\\
{\mathcal C}\ii\Theta \iiijm +{\mathcal C}\jj\Theta \iimjj+\\
+({\mathcal C}\ii-{F})\Theta\ii+({\mathcal C}\jj-F)\Theta\jj=0
\end{array}
\end{equation}
then the function
\[\Phi=\Theta\ii+ \Theta\jj\]
satisfies the equation formally adjoint to eq. \mref{10}
\begin{equation}
\begin{array}{l}
({\mathcal A}+{\mathcal B}) \Phi + ({\mathcal A}\jjm +{\mathcal B}\iim ) \Phi\jimjm+
{\mathcal A}\jijm\Phi\jijm+\\ {\mathcal B}\jimj\Phi\jimj
+{\mathcal A}\iim \Phi\iimim+{\mathcal B}\jjm \Phi\jjmjm+ \\
{\mathcal C}\jimj\Phi \iimimj +{\mathcal C}\jijm\Phi \iijmjm + \\
+({\mathcal C}-{F}\iim )\Phi\iim +({\mathcal C}-F\jjm )\Phi\jjm =0
\end{array}
\end{equation}
The scheme is proper for solving the initial-boundary value problem we discussed in section
\ref{boundary}.

We finally receive transformation
\begin{equation}
\begin{array}{l}
\frac{\Dj (( \bar{\Psi}\ii+\bar{\Psi}\jj)(\Theta\ii+ \Theta\jj))}{(\Theta + \Theta\jimj)}= \\
  \left[{\mathcal A} (\Theta\ii + \Theta\jj)  +  
                           {\mathcal C} (\Theta + \Theta\jijm)\right] 
  \Dim \frac{{\Psi}\ii+{\Psi}\jj}{{\Theta}\ii+{\Theta}\jj}- \\
\hfill{ {\mathcal C}  (\Theta+ \Theta\jijm) \Djm \frac{{\Psi}\ii+{\Psi}\jj}{{\Theta}\ii+{\Theta}\jj}}
\\
\\
\frac{\Di (( \bar{\Psi}\ii+\bar{\Psi}\jj)(\Theta\ii+ \Theta\jj))}{(\Theta + \Theta\jijm)}= \\
- \left[ {\mathcal B}  (\Theta\ii+ \Theta\jj)   +
                           {\mathcal C}  (\Theta + \Theta\jimj)\right] 
\Djm \frac{{\Psi}\ii+{\Psi}\jj}{{\Theta}\ii+{\Theta}\jj}-\\
\hfill{ {\mathcal C} (\Theta+ \Theta\jimj) 
\Dim \frac{{\Psi}\ii+{\Psi}\jj}{{\Theta}\ii+{\Theta}\jj}}
\end{array}
\end{equation}

\section{Conclusions and perspectives}
\label{discuss}
We end the paper with some conclusions and propositions for further development.

The generalization of results of this paper 
to multi-dimension can be given where set of difference operators
$\Delta_i, \, i=1,...,N$ is replaced by a set operators $D_i, \, i=1,...,N$ that are linear commuting combinations of $\Delta_i$.
The same reasoning holds as far as Laplace transformations are concerned.
The 6-point scheme admit factorization $[(M_1 T_1+N_1 T_2+X_1) (M_2 T_1+N_2 T_2+X_2)+H] \psi=0$
(where $T_1$ and $T_2$ are shifts in $m$ and $n$ direction respectively.
so it can be used to develop theory of Laplace transformations
for difference equations
\cite{Dol,Novikov,NovDyn,AdlSta,NieL,OblPen}.

We would like to mention that on this level of generality 
showing $q$-difference analogue discrete schemes we just introduced
is straightforward \cite{PMalk}.

Four propositions of further developments are in order.
First, generalization of quadratic  and symmetric reductions to 6-point 
scheme (it will be given in forthcoming papers).
Second, simple idea  by professor
Decio Levi that not only operators
$\Di$ $\Dj$ $\Dim$ $\Djm$ can be of importance e.g. one can try to use $T_1-T_2$ and  $T_1T_2-1$
operators instead.
Third, to investigate the role of just presented transformations in the difference geometry.
Finally and the most importantly,  deriving  hierarchies of nonlinear equations associated with all the equations
presented in the paper.

Since we concentrated here on discretizations of hyperbolic equations we end the article with three comments on
star schemes -- proper discretizations of elliptic differential equations.
Firstly, let us notice that if we substitute
$\psi= \phi+ \phi \ii+ \phi \jj$ into the 6-point scheme we will obtain star-like difference scheme, which
can serve as star like discretization of eq. \mref{f} and which needs further studies.

Secondly, it is remarkable that star like (or cross like) operators such as 7-point scheme or 5-point scheme
appeared in the integrable literature occasionally \cite{Kri2,Novikov,Frank,NovDyn,Frank2,OblPen}.
But almost none of the results were used to obtain solutions of nonlinear integrable systems.
The only exceptions are works concerning discrete time Toda chains  
\cite{Hir,Hir1,Sur,Sur1,Sur2,Sur3}
which are a nonlinear 5-point scheme
itself and the work \cite{NSD} the result of which were used to obtain solutions of generalization of Toda chain
to two dimensional lattice \cite{SND}. 
In other words theory of integrable difference systems
based on the schemes other than 4-point schemes is still in its infancy.

Finally, the relationship of integrable systems on quad-graphs with equations on stars  
the  discrete time Toda type lattices 
are  established \cite{Adler,BS,BS2,ABS} and we would like to refer to the relationship as to sub-lattice approach. 
It is not clear under what circumstances 
the sub-lattice approach  does not destroy integrability.
However in the paper \cite{DGNS} it was proved that for discrete lattices governed by Moutard equation
integrability features like existence of Darboux transformations are inherited by a sub-lattice.

{ \bf Acknowledgments }
The main part of the paper has been presented as a poster during the conference  Symmetries and Integrability of Difference Equations SIDE VI June 19-24, 2004 Helsinki.
I would like to express my thankfulness to organizers of the  SIDE VI conference
for the support that enable me to take part in the fruitful conference.
 I would like to acknowledge    
Paolo Maria Santini for motivating me to write the paper, 
Jan Cie\'sli\'nski for indicating me that Combescure transformation is fundamental in Jonas's
fundamental transformation and 
Frank Nijhoff for the literature guidance.

\end{document}